%% file: inoculation.tex
\documentclass[sigconf]{acmart}

\makeatletter
\renewcommand\addcontentsline[3]{%
}
\makeatother

\usepackage{subcaption}

\usepackage{enumitem}
\usepackage{array}
\usepackage{float}
\usepackage{tabularray}
\usepackage{tabularx}
\usepackage{booktabs}
\usepackage{svg}
\usepackage{multirow}
 
\usepackage{fvextra}

\newlist{researchquestions}{enumerate}{1} 
\setlist[researchquestions]{label=\textbf{RQ\arabic*}, noitemsep, topsep=-0.01pt}

\AtBeginDocument{%
  }

\setcopyright{acmlicensed}
\copyrightyear{2025}
\acmYear{2025}
\setcopyright{rightsretained}
\acmConference[CHI EA '25]{Extended Abstracts of the CHI Conference on Human Factors in Computing Systems}{April 26-May 1, 2025}{Yokohama, Japan}
\acmBooktitle{Extended Abstracts of the CHI Conference on Human Factors in Computing Systems (CHI EA '25), April 26-May 1, 2025, Yokohama, Japan}\acmDOI{10.1145/3706599.3720125}
\acmISBN{979-8-4007-1395-8/2025/04}

\begin{document}

\title[Effective Yet Ephemeral Propaganda Defense]{Effective Yet Ephemeral Propaganda Defense: There Needs to Be More than One-Shot Inoculation to Enhance Critical Thinking}

\author{Nicolas Hoferer}
\email{nicolas.hoferer@uzh.ch}
\affiliation{%
  \institution{University of Zurich}
  \city{Zurich}
  \country{Switzerland}
}

\author{Kilian Sprenkamp}
\email{kilian.sprenkamp@uzh.ch}
\affiliation{%
  \institution{University of Zurich}
  \city{Zurich}
  \country{Switzerland}
}

\author{Dorian Christoph Quelle}
\email{dorian.quelle@math.uzh.ch}
\affiliation{%
  \institution{University of Zurich}
  \streetaddress{}
  \city{Zurich}
  \country{Switzerland}
}

\author{Daniel Gordon Jones}
\email{danielgordon.jones@uzh.ch}
\affiliation{%
  \institution{University of Zurich}
  \city{Zurich}
  \country{Switzerland}
}

\author{Zoya Katashinskaya}
\email{zoya.katashinskaya@uzh.ch}
\affiliation{%
  \institution{University of Zurich}
  \city{Zurich}
  \country{Switzerland}
}

\author{Alexandre Bovet}
\email{alexandre.bovet@uzh.ch}
\affiliation{%
  \institution{University of Zurich}
  \streetaddress{}
  \city{Zurich}
  \country{Switzerland}
}

\author{Liudmila Zavolokina}
\email{liudmila.zavolokina@unil.ch}
\affiliation{%
  \institution{University of Lausanne}
  \streetaddress{}
  \city{Lausanne}
  \country{Switzerland}
}
\affiliation{%
  \institution{University of Zurich}
  \city{Zurich}
  \country{Switzerland}
}

\renewcommand{\shortauthors}{Hoferer et al.}

\input{sections/abstract}



\received{23 January 2025}
\received[revised]{06 March 2025}

\maketitle
\setlength{\textfloatsep}{6pt}     
\setlength{\intextsep}{6pt}        
\setlength{\floatsep}{6pt}  
\input{sections/01_intro}
\input{sections/02_related_work}
\input{sections/03_tool_design}
\input{sections/04_method}
\input{sections/05_findings}
\input{sections/06_discussion_conclusion}

\begin{acks}
We thank the Digital Society Initiative of the
University of Zurich and the Digitalization Initiative of the Zurich
Higher Education Institutions (DIZH) for financing this
study under the DIZH Founder Call.
\end{acks}

\bibliographystyle{ACM-Reference-Format}
\bibliography{references}

\input{sections/appendix}

\end{document}

%% file: sections/abstract.tex
\begin{abstract}
In today's media landscape, propaganda distribution has a significant impact on society. It sows confusion, undermines democratic processes, and leads to increasingly difficult decision-making for news readers.
We investigate the lasting effect on critical thinking and propaganda awareness on them when using a propaganda detection and contextualization tool.
Building on inoculation theory, which suggests that preemptively exposing individuals to weakened forms of propaganda can improve their resilience against it, we integrate Kahneman's dual-system theory to measure the tools' impact on critical thinking. Through a two-phase online experiment, we measure the effect of several inoculation doses.
Our findings show that while the tool increases critical thinking during its use, this increase vanishes without access to the tool.
This indicates a single use of the tool does not create a lasting impact. We discuss the implications and propose possible approaches to improve the resilience against propaganda in the long-term.
\end{abstract}

%% file: sections/01_intro.tex
\section{Introduction}
According to the World Economic Forum's Global Risks Report 2024 \cite{mclennan2024global}, spreading misinformation and propaganda is a major global risk. It can undermine democratic values, increase societal polarization and lead to violence \cite{wefglobalrisks2024}.
The rise of generative AI, especially Large Language Models (LLMs), has simplified rapid propaganda distribution, while the difficulty of distinguishing between fake and real news complicates combating its spread.

Inoculation theory \cite{Pfau1988INOCULATIONIP, Banas2020InoculationT} offers a promising approach for building resistance against propaganda. 
By exposing individuals to a weakened form of propaganda, it strengthens resistance to future, more potent attempts \cite{dillard_sage_2024}. It works analogously to medical vaccines, strengthening individuals' ``immunity'' against propaganda.
For instance, Roozenbeek \textit{et al.} \cite{roozenbeek_psychological_2022} demonstrated that short inoculation videos improved resilience against propaganda on social media, though their impact on critical thinking (CT) remains unclear.

While traditional inoculation approaches rely on human-designed interventions, technological developments offer new possibilities for implementing them at scale. Machine learning approaches have recently been developed for the task of propaganda detection \cite{DellaVedova8468301, vorakitphan:hal-03417019}, including LLMs-based methods \cite{sprenkamp2023large}. Automatic detection of propaganda enables the creation of tools that nudge users about propaganda techniques used in online news articles they are reading. For example, Zavolokina \textit{et al.} \cite{zavolokina_think_2024} demonstrated that an LLM-based propaganda detection tool enhances CT and propaganda awareness through nudging.
While this work highlights the immediate improvements in users’ CT and awareness by such tools, the question remains whether these effects persist once the tool is no longer used.

To address this gap, we developed a browser extension for propaganda detection and contextualization. Our tool implements inoculation theory through three key functions: detection and highlighting of potential propaganda, explanations for deconstructing propaganda techniques, and presentation of additional contextual information. The contextual information is curated factual information and source details that help users understand why a statement may be propagandistic and how it fits into the broader context. Together, these features expose users to propaganda in a controlled way while simultaneously providing them with the understanding and tools to avoid being influenced by it - the core principles of inoculation theory.

We design five levels of exposure: (1) a control group receiving no inoculation intervention, (2) a low-dose group only receiving an overview of propaganda techniques, (3) a first medium-dose group using the tool with propaganda detection and explanations, (4) a second medium-dose group getting the detection and contextual information related to the statement but no explanation, and (5) a high-dose group receiving the detection, the explanations and the contextual information.
The aim of these varying levels of exposure is to see if the inoculation doses affect users' CT and propaganda awareness differently.
Our work specifically aims to investigate the following two research questions:

\begin{researchquestions}
\item Does exposure to a propaganda detection and contextualization tool lead to lasting improvements in CT after the tool is no longer available?
\item How do different inoculation doses (low, medium, high) affect CT and propaganda awareness?
\end{researchquestions}

Our findings indicate that while the tool improves CT and propaganda awareness during its use, these improvements do not last once the tool is no longer accessible. Even participants who showed a positive immediate effect returned to baseline levels without the tool. These results suggest our approach is not sufficient to achieve long-term improvement. We conclude our work by proposing alternative approaches.

%% file: sections/02_related_work.tex
\section{Related work}
The challenge of combating propaganda effectively requires an approach that provides immediate help but also builds long-term resistance.
Although propaganda detection systems \cite{DellaVedova8468301, sprenkamp2023large} can provide immediate assistance, their limited accessibility shows the need for publicly available solutions that also help news readers develop independent CT \cite{sprenkamp2023large}.

\subsection{Propaganda Detection and Contextualization}
Propaganda involves intentionally manipulating opinions using rhetorical and psychological strategies, such as loaded language (i.e., employing words or phrases with strong emotional implications) \cite{DaSanMartino2020Prta}. Traditional approaches framed propaganda detection as a classification problem \cite{martino2020semeval, alam2022overview, al2019justdeep, da2019fine}, requiring extensive labeled datasets and offering only limited interpretability.
Da San Martino \textit{et al.} \cite{DaSanMartino2020Prta} systematically analyzed and defined 14 propaganda techniques, providing a comprehensive framework for understanding propaganda (detailed in Table \ref{Table: Propaganda Techniques} in the Appendix \ref{app:propaganda_techniques}). We leverage their established taxonomy of propaganda techniques as a basis for our tool.

Recent developments in LLMs opened up new possibilities \cite{sprenkamp2023large}. In particular, LLMs offer three key advantages. First, they enable propaganda detection through prompt-based learning without the need for specialized datasets. Secondly, they can generate explanations about why it may be propaganda \cite{zavolokina_think_2024}, making it more transparent and educational for users.
Third, they are able to retrieve and synthesize contextual information, helping users evaluate content better.

Several approaches have been developed, from automated fact-checking to generalist information retrieval systems \cite{quelle2024perils, wang2023factcheck, tang2024minicheck, zhang2023towards}. While fact-checking systems show potential, their discrete true/false classifications can lead to excess false positives when dealing with predominantly true information. Instead of solely relying on LLM judgments, a more effective approach may be to use LLMs to retrieve and synthesize relevant contextual information, allowing readers to form their own informed conclusions.

Our tool combines three components: first, propaganda is detected, then the tool provides both an explanation of the techniques used and relevant contextual information to help readers evaluate the content. This integrative approach aims not only to nudge readers to potential propaganda but also to support their understanding of how propaganda works and provide context for independent assessment, which should lead to improved CT.

\subsection{Critical Thinking and Inoculation Theory}
Our approach builds on two theoretical frameworks: dual-process theory and inoculation theory.
CT refers to a human's ability to analyze facts, evidence, and arguments to form a judgment through rational, skeptical, and unbiased evaluation \cite{glaser2017defining}. Dual process theories explain how we think and make decisions by categorizing cognitive processes into two systems: System 1 (automatic) and System 2 (reflective) \cite{kahneman2011thinking, strack2004reflective}. System 1 is fast and intuitive, handling routine tasks like driving, while System 2 is slower and more deliberate, allowing for in-depth analysis \cite{kahneman2011thinking}. Though System 2’s thoughtful approach seems superior, System 1’s efficiency in managing everyday tasks—like walking or talking—is crucial. 
When dealing with multiple demanding tasks, people may switch to System 1, even when System 2’s critical analysis would be more suitable. This shift can lead to reliance on heuristics, which can introduce biases like confirmation bias, favoring information that supports existing beliefs \cite{caraban201923}.
Kahneman’s studies identify six key characteristics of dual process thinking, summarized in Table \ref{tab:dual_system_thinking} in the Appendix \ref{app:CT_characteristics}. These are essential for understanding how individuals engage in CT and whether they rely on System 1 or System 2 for decisions. We use these characteristics in our study to capture the CT of our participants, i.e., news readers. Despite the broad acceptance of dual process theory, it has been criticized for oversimplifying cognitive processes by dividing them into two distinct systems \cite{evans2013dual}. 

Inoculation theory \cite{Pfau1988INOCULATIONIP} offers a relevant extension to dual process thinking, particularly in the context of resistance to propaganda. Here, the idea is to strengthen resistance to propaganda by exposing individuals to weakened forms of deceptive content or to refute it \cite{Compton06072024}.
This approach has been successfully applied by Maertens \textit{et al.} \cite{maertens_long-term_2021}. They investigated the long-term effectiveness of inoculation against misinformation. Recent work used inoculation with conversational agents and LLMs to improve propaganda awareness \cite{russo2023aide}. Roozenbeek \textit{et al.} used inoculation videos \cite{roozenbeek_psychological_2022} and games \cite{roozenbeek2019fake} to improve propaganda awareness.
Furthermore, Bastani \textit{et al.} \cite{Bastani2024} showed how LLMs can hinder skill acquisition and even harm long-term learning, even when there might be an initial improvement in immediate performance. 
Similarly, Kasneci \textit{et al.} \cite{KASNECI2023102274} observe that generative AI tools may be used as a "crutch", thus diminishing users' problem-solving skills.

To our best knowledge, no previous studies have combined these two theories within the context of propaganda detection and contextualization tools and different inoculation dosages. While existing work showed immediate benefits, it remains unclear how long inoculation effects persist in terms of CT. This work offers an opportunity for designing interventions that promote CT and improve resistance to propaganda, particularly in times of increasingly realistic manipulative content.

%% file: sections/03_tool_design.tex
\section{Propaganda Detection and Contextualization Tool}
Our tool is a browser extension designed to assist users in identifying propaganda within news articles. Figure \ref{fig:usageExample} shows an example usage of the tool. The architecture diagram is shown in Appendix \ref{app:architecture}.

\begin{figure*}[t]
    \centering
    \includegraphics[width=0.71\textwidth]{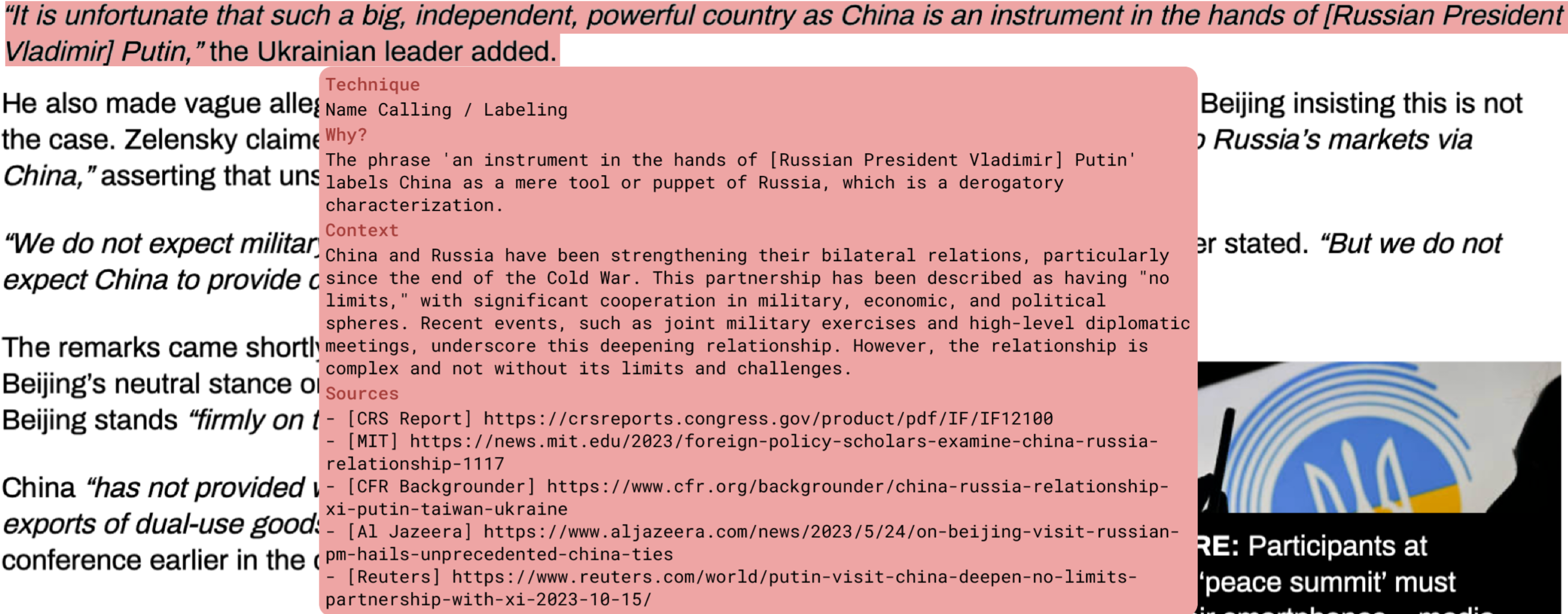}
    \vspace{-10pt}
    \caption{Example Usage}
    \label{fig:usageExample}
\end{figure*}

A user can select text on a webpage via a right mouse click and choose propaganda detection with or without contextualization. The marked text is sent to a FastAPI endpoint where a prompt template (Appendix \ref{app:propaganda-detection-prompt-template}) instructs GPT-4o \cite{openai2024gpt4ocard} to detect propaganda. The OpenAI API \cite{openai_api} identifies propaganda from the 14 techniques defined by Da San Martino \textit{et al.} \cite{da-san-martino-etal-2020-semeval} and provides explanations. The frontend highlights the propaganda and adds annotations, as shown in Figure \ref{fig:usageAndArchitecture}, by matching the output to the the correct locations on the website.

The ``contextualizer'' provides the user with factual information on a selected text using Google Cloud Platform's search capabilities with specialized source filtering mechanisms. The filtering process excludes sources through a predefined database of problematic websites scraped from Media Bias/Fact Check (MBFC)\footnote{https://mediabiasfactcheck.com} that includes all links classified as ``questionable''. This filtering helps ensure that when users search for text from an article, they are not redirected back to unreliable sources that may have originated the misinformation.

The response format has two main sections: Context and Sources. The Context section presents simplified, factual information about the topic, while the Sources section lists the used references. This structure helps users verify information and build resistance to propaganda by making facts and their verification easily accessible. The contextual information is generated and extracted from the search result by an LLM using the Reasoning and Act (ReAct) prompt framework \cite{yao2023reactsynergizingreasoningacting}.
The prompt template is available in Appendix \ref{app:contextualizer-propmpt-template}.

%% file: sections/04_method.tex
\section{Methodology}
We conducted an online experiment to investigate whether exposure to our tool could lead to lasting improvements in CT and propaganda awareness. Specifically, the experiment was structured into two phases, each separated by one week, to capture immediate and short-term effects on participants' CT and propaganda awareness. In the experiment, participants were first introduced to the tool, given access to it, and then asked to complete surveys that captured their perceptions. This design systematically controlled the exposure while measuring the effects consistently.

Participants were recruited through Prolific \cite{prolific}, an online platform commonly used in academic research because of its quality and reliability \cite{EyalProlificDataQuality}. The compensation was set to the platform's recommended rate of 9 GBP / hour.
To ensure a diverse yet controlled participant pool, we applied the following selection criteria: participants had to be native English speakers, have no history of language disorders, and possess no background in media or political studies. These requirements ensured that language comprehension was not a barrier and participants lacked prior expertise in the study’s subject matter. 
We used several news articles to evaluate the impact on CT and propaganda awareness.
The articles varied in topic and content but were comparable in size and complexity, as given by the Flesch-Kincaid
Grade Level (Article 1: 12.0, Article 2: 12.3, Article 3: 12.2) \cite{FleschKincaid}.

Participants were randomly assigned to one of five groups exposed to a different dosage of inoculation, as shown in Table \ref{tab:group-overview}.
For the pretest, we recruited five participants per group. The total number of participants per group is listed in Table \ref{tab:group-overview}. For Survey 2, we aimed to retain the same participants as in Survey 1, but some dropped out, leading to lower participation numbers.
\begin{table*}[t]
    \centering
    \begin{tabular}{lcccp{6.5cm}}
        \textbf{Group} & \textbf{Dosage} & \textbf{Survey 1} & \textbf{Survey 2} & \textbf{Description} \\
        \midrule
        Group 1 & None (control) & 100 & 82 & Read Article 2 without any intervention \\
        \addlinespace[0.4em]
        Group 2 & Low & 103 & 86 & Overview of 14 propaganda techniques (explanations and examples from \cite{da-san-martino-etal-2020-semeval}, see Appendix \ref{app:propaganda-detection-prompt-template}), then read Article 2. \\
        \addlinespace[0.4em]
        Group 3 & Medium I & 102 & 81 & Tool (detected techniques, and explanations) for Article 1, then read Article 2 without tool \\
        \addlinespace[0.4em]
        Group 4 & Medium II & 101 & 87 & Tool (detected techniques, and context) for Article 1, then Article 2 without tool \\
        \addlinespace[0.4em]
        Group 5 & High & 100 & 88 & Tool (detected techniques, explanations and related context) for Article 1, then Article 2 without tool \\
    \end{tabular}
    \caption{Experimental Groups and Participation Overview}
    \label{tab:group-overview}
\end{table*}

Survey 1 measured the immediate effect, and participants completed the following three steps. First, depending on their group, they read the assigned article(s) with or without using our tool. Next, participants completed standardized questionnaires to evaluate their CT ability and propaganda awareness. Finally, groups that used the tool (Groups 3-5) provided feedback on their experience and the perceived quality of the tool. Participants from Group 2 provided feedback on the 14 propaganda techniques.

One week after Survey 1, we conducted Survey 2 to assess how well participants retained CT and propaganda awareness.
All participants read the same article (\textit{Article 3}) without using the tool, then completed the same standardized questionnaires as before to assess retention.

We employ a set of measurements to evaluate participants' CT abilities, propaganda awareness, and overall experience with the tool, such as the Need for Cognition Scale \cite{needforcognition}, a psychologically grounded approach to assess the thinking tendencies of a participant.  For comparability, we adopted the measurement set from Zavolokina et al. \cite{zavolokina_think_2024} (see Appendix \ref{app:measurements}).

Before reading the assigned articles, participants filled out a questionnaire to assess their political leanings, news consumption habits, and trust in the tool as well as the news article.
After reading each assigned article, participants were asked to complete a series of post-reading questionnaires. To measure propaganda awareness, participants reported whether they believed the article contained propagandistic elements and explained their reasoning. Then, they evaluated specific statements to determine whether they were propagandistic.

%% file: sections/05_findings.tex
\section{Findings}
\subsection{Survey 1 - Post Treatment Effect on CT and Propaganda Awareness}
Survey 1 measures participants' CT and propaganda awareness in two phases. The first phase is during the inoculation treatment (i.e., with the overview for group 2 and tool assistance for groups 3 to 5), and the second phase is after the inoculation treatment when reading a new article without the overview or tool assistance.

\begin{figure*}[htbp]
    \centering
    \includegraphics[width=\textwidth]{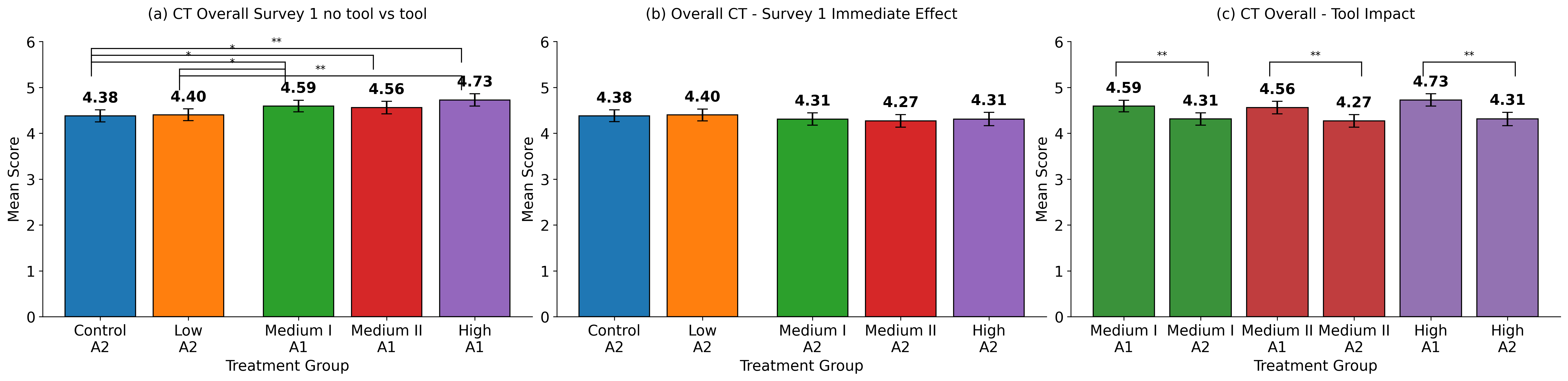}
    \vspace{-15pt}
    \caption{CT assessment across different conditions: (a) Comparison of CT between control/low-dose groups (reading Article 2 without tool) vs. medium/high-dose groups (reading Article 1 with tool), showing significant enhancement when using the tool; (b) Post-inoculation CT when all groups read Article 2 without tool access, demonstrating no sustained improvement; (c) Within-group comparison showing significant decline in CT when tool access was removed. Error bars represent standard error. Statistical significances: * p < 0.050 ** p < 0.010}
    \label{fig:critical_thinking_results}
\end{figure*}

In the initial phase, we evaluate the tool's effectiveness in increasing CT and propaganda awareness by replicating Zavolokina \textit{et al.}'s experiment \cite{zavolokina_think_2024}. We compare the control and low-dosage groups (who read Article 2 without the tool) against the medium and high-dosage groups (who read Article 1 with the tool). Even though the articles are different, this comparison allows us to assess the immediate effect of the tool. Our findings indicate that participants with access to the tool demonstrate a significant increase in CT and propaganda awareness compared to those without it. 
Specifically, $t$-tests between Group 1 (control) and Groups 3, 4 and 5 show significant differences (1 vs. 3: $t = -2.329$, $p = 0.020$; 1 vs. 4: $t = -2.001$, $p = 0.048$; and 1 vs. 5: $t = -3.693$, $p = 0.000$, respectively), as does the comparison between Group 2 (low-dosage) and Groups 3 and 5 (2 vs. 3: $t = -2.078$, $p = 0.038$; and 2 vs. 5: $t = -3.460, p = 0.001$, respectively). $p$-values between the other groups are all above 0.083. These results align with the previous research \cite{zavolokina_think_2024} and confirm that such a tool increases CT during its use (see Figure \ref{fig:critical_thinking_results} (a)).

However, to answer our research question about whether these improvements persist without access to the tool, we examine participants' performance on article 2, i.e. without access to the tool or the technique overview.
Independent $t$-tests reveal no statistically significant difference in overall CT between the groups regardless of treatment dosage when reading article 2, as shown in Figure \ref{fig:critical_thinking_results} (b) (all $p$-values are higher than 0.150).

In addition, we examine the effect for the medium and high dosage groups when transitioning from using the tool while reading article 1 to not using it when reading article 2.
We use dependent $t$-tests and observe that as soon as participants do not have access to the tool, their CT drops statistically significantly (see Figure \ref{fig:critical_thinking_results} (c)).
Specifically, the $p$-value for each group with vs. without the tool is less than 0.001.
This indicates that there is no sustained transfer effect after using the tool once, and participants are not able to maintain the same level of CT as during the use of the tool. 

\begin{figure*}[htbp]
    \centering
    \includegraphics[width=0.9\textwidth]{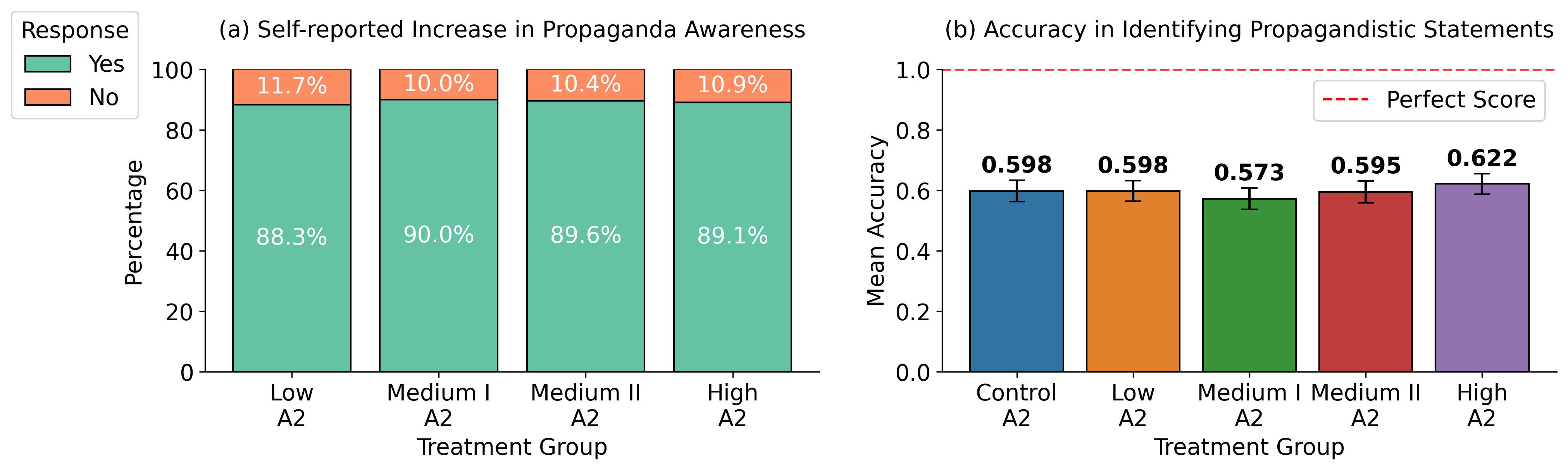}
    \vspace{-10pt}
    \caption{Results on propaganda awareness showing (a) self-reported increase in propaganda awareness across treatment groups, demonstrating higher perceived awareness in groups that used the tool and (b) actual accuracy in identifying propagandistic statements, showing no significant difference between groups despite varying treatments. Error bars in (b) represent the standard error.}
    \label{fig:propaganda_awareness}
\end{figure*}

Regarding propaganda awareness, we observe a disconnect between self-reported scores and the demonstrated ability to detect propaganda. Participants with access to the tool report feeling more aware of propaganda (Figure \ref{fig:propaganda_awareness}). However, their actual performance in identifying propagandistic statements does not show statistically significant differences. 
This suggests that while participants are more aware of propaganda when supported by the tool, this increased awareness does not carry over to their independent analysis of new content.

\subsection{Survey 2 - One-Week Retention}
In Survey 2, we evaluate the retention of CT and propaganda awareness without access to the tool for all groups. The participants are all presented with the same new article and complete identical questionnaires to ensure comparability.
We use dependent $t$-tests to compare CT scores within each group between the two surveys and independent $t$-tests to compare performance between groups within the second survey.

The results show no statistically significant differences in CT scores within groups between the surveys (minimum $p = 0.015$ for Group 4, $t = -2.447$). All groups have similar CT scores to those in the first survey.

We do not find consistent statistically significant differences in CT between survey 2 groups, with only one comparison showing significance: Group 1 vs. Group 4 ($t = -2.141$, $p = 0.033$). All other group comparisons are not statistically significant (all $p > 0.110$), suggesting broadly similar CT levels across groups in the second survey.

This finding suggests that the inoculation does not lead to lasting improvements in CT, no matter the dosage. The results show that a single exposure to the tool or the propaganda technique overview is insufficient to create lasting improvements in CT one week after the treatment.

%% file: sections/06_discussion_conclusion.tex
\section{Discussion and Conclusion}
Our work investigated whether and how an AI-based propaganda tool, designed with elements of inoculation, could foster CT in news readers. Although our initial formulation suggested directly applying inoculation theory, our findings reveal a more nuanced relationship between technological assistance and inoculation principles.

Our approach integrated weakened forms of propaganda through three mechanisms: detection of propaganda, explanations, and contextual information. We provide empirical evidence of how these technological mechanisms and psychological frameworks interact. In our experimental setup, participants were exposed to propaganda content alongside the tool's annotations, effectively creating a form of weakened exposure that aligns with inoculation theory principles. However, in practical applications, the tool functions differently. Rather than pre-exposing users to weakened forms of propaganda, it provides real-time analysis and annotations that help "weaken" propaganda techniques as users encounter them naturally on websites.

A key finding is the temporal nature of the tool's intervention. The core challenge appears to be that participants used the tool as a "crutch" rather than developing independent CT. This dependency aligns with observations from recent research on LLM-based tools \cite{Bastani2024}, where overreliance also happened. While participants had access to the tool, their CT and propaganda awareness increased significantly. However, without access to the tool, these increases vanished. This tool dependency suggests users defaulted to System 1 thinking instead of developing the deeper, more deliberate System 2 thinking processes essential for genuine CT. Furthermore, variations in engagement levels across experimental conditions may have influenced these outcomes, i.e., through the using the tool, participants automatically interacted more with a given article and thus thought more critically. While this is a desired quality for the tool, the experimental setup should be adjusted for better comparability between treatments.

Our experiments confirm a direct effect of the tool on CT and propaganda awareness, aligning with the findings of Zavolokina et al. \cite{zavolokina_think_2024}. During its use, participants demonstrated significantly improved CT, indicating that the tool effectively supports users in analyzing propaganda in real-time. However, this effect did not persist beyond immediate usage. Both directly after using the tool and in the follow-up assessment one week later, participants showed no significant difference compared to treatments without the tool, showing no evidence of inoculation. These results highlight the challenge of achieving lasting improvements in CT through a one-shot intervention and emphasize the need for repeated exposure or alternative educational strategies to reinforce independent CT skills over time.

When comparing our approach to successful inoculation implementations, particularly Roozenbeek \textit{et al.'s} \cite{roozenbeek_psychological_2022} work, some key differences in implementation and mechanisms emerge. They used short videos, engaging stories, and visual elements to expose participants to weakened forms of propaganda techniques. In contrast, our tool attempted to weaken existing propaganda through annotations. This suggests that our implementation of inoculation might be an issue and that the mode of delivery plays a significant role.

Our evaluation using Kahneman's framework extends beyond previous work, such as Russo \textit{et al.'s} \cite{russo2023aide} conversational agent-based study. While their work demonstrated benefits through story creation and evaluation, our approach revealed a crucial gap between assisted and independent CT. This measurement approach allowed us to quantify not just the immediate effects of AI assistance but also explain the cognitive mechanisms behind tool dependency.

Several other factors might explain the limited retention of CT toward propaganda. First, the lack of emotional elements in our approach could play an important factor, as emotional engagement has been shown to play a crucial role in learning retention \cite{PHELPS2004198}. Second, as Maertens \textit{et al.} \cite{maertens_long-term_2021} found in their work around the long-term effectiveness of inoculation, a single exposure might be insufficient, and repeated treatments or "boosters" might be required. Third, our measurement methodology for CT relied on self-reported questionnaires, and the study setup might not be optimally suited for testing inoculation effects, given the lack of participants' motivation to improve CT. Here, the choice of Prolific as a recruitment platform may have influenced our results, as participants may not have had a strong intrinsic motivation to develop lasting resistance to propaganda. Unlike real-world users who actively seek to reduce their exposure to manipulative content, Prolific participants engage in studies for compensation rather than personal interest in consuming less propaganda. Complementary to self-reporting, alternative approaches such as eye-tracking, neurological studies, or longitudinal observational studies could provide deeper insights into whether actual inoculation effects occur. Our study also raises important considerations about the content generated by LLM-based tools. While our tool aims to detect and explain propaganda and provide contextual information, it could also introduce its own biases or misinformation.

These insights suggest several promising directions for future research and development. Future research should explore interactive, emotional, and multimedia-based interventions to make the learning process more compelling and address engagement challenges. However, future experiments should be set up in such a manner that different treatment groups require the same engagement level, a shortcoming of our study. Additionally, further studies should not solely rely on self-reported metrics but include metrics such as eye-tracking, neurological studies, or longitudinal observation. Further, the tool could be improved pedagogically to avoid being used as a crutch, perhaps by gradually reducing the level of annotation support while maintaining its ability to weaken propaganda techniques in real-time. Additionally, the issue of sustaining learned skills could be tackled through spaced repetition or booster interventions that combine the tool's real-time weakening capabilities with more traditional inoculation approaches. Future implementations could also incorporate storytelling elements, interactive exercises, or gamification to actively engage users in identifying propaganda techniques and to activate CT.

Combating propaganda requires bridging the gap between immediate effect and long-term improvements in CT and propaganda awareness. One path forward could lie in creating a system that better balances real-time propaganda weakening with lasting skill development, focusing on strengthening users' ability to think critically and independently about the information they read while carefully considering the potential risks of using AI-generated content. Overall, our findings show that, while AI-based tools can effectively help with CT-heavy tasks, creating lasting improvements requires a more sophisticated approach that combines technological assistance with established psychological principles of learning and behavior change.

%% file: sections/appendix.tex
\appendix
\section*{APPENDIX}
\addcontentsline{toc}{section}{APPENDIX}
\section{Tool Architecture and Usage}
\label{app:architecture}
\begin{figure*}[!ht]
    \centering
    \includegraphics[width=\textwidth]{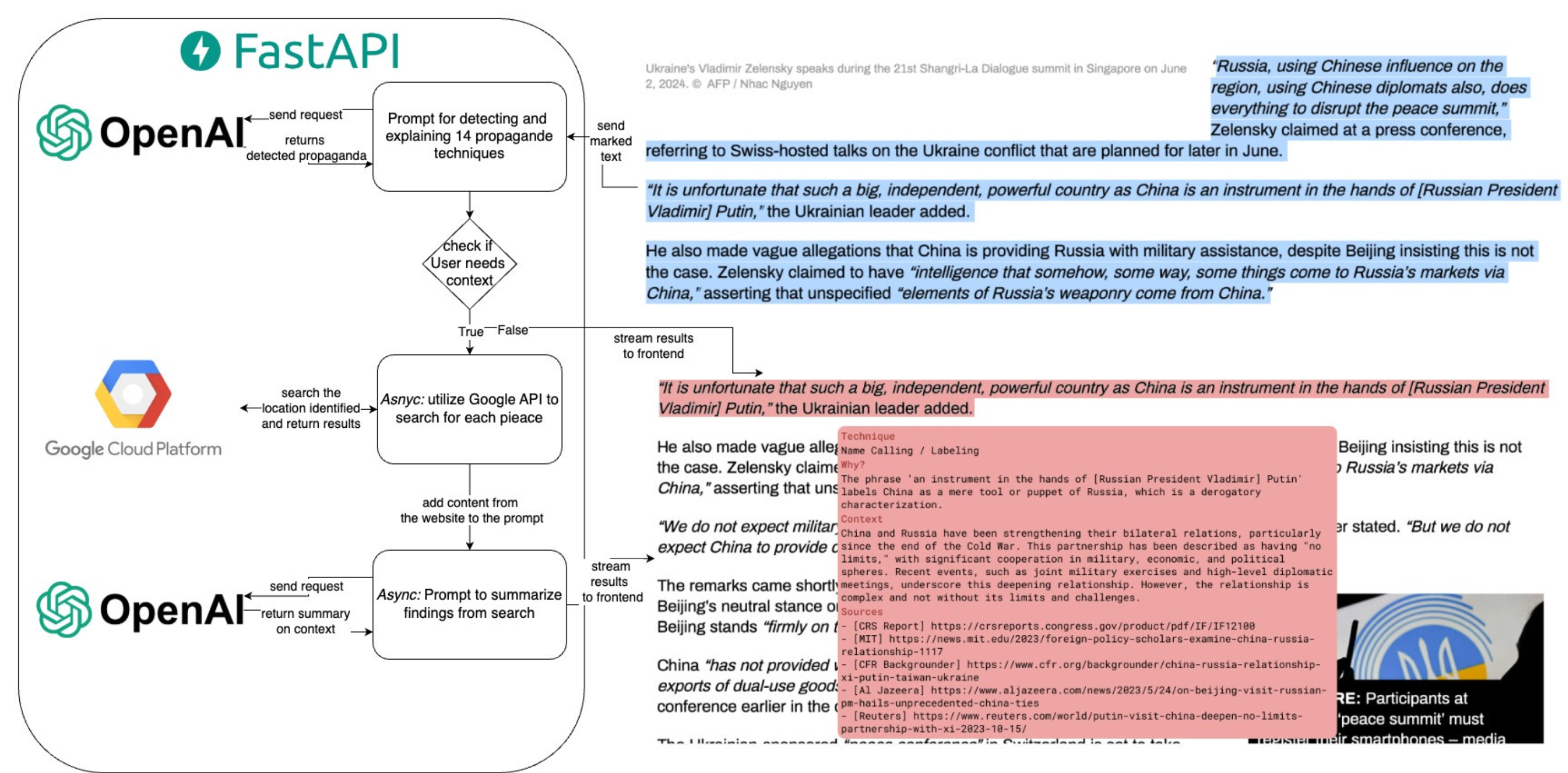}
    \caption{Example Usage and Architecture}
    \label{fig:usageAndArchitecture}
\end{figure*}

Figure~\ref{fig:usageAndArchitecture} shows the tool's architecture and how it works.

\section{Critical Thinking Characteristics}
\label{app:CT_characteristics}
Table~\ref{tab:dual_system_thinking} shows the key characteristics of dual system thinking based on Kahneman's work. It contrasts System~1 (fast, automatic, and effortless) with System~2 (slow, controlled, and effortful) across six characteristics (speed, processing, control, effort, nature, and adaptability).

\begin{table*}[!ht]
    \centering
    \caption{Dual system thinking characteristics extracted from \cite{MapsOfBoundedRationality, TwoSystemsInTheMind, kahneman2011thinking}}
    \label{tab:dual_system_thinking}
    \begin{tabular}{p{3cm}p{5cm}p{3cm}p{3cm}}
        \textbf{Characteristic} & \textbf{Definition} & \textbf{System~1} & \textbf{System~2} \\
        \midrule
        Speed         & Speed of thinking                 & Fast          & Slow         \\
        Processing    & Approach to handling thinking tasks & Parallel      & Serial       \\
        Control       & Degree of conscious oversight     & Automatic     & Controlled    \\
        Effort        & Cognitive load                    & Effortless    & Effortful     \\
        Nature        & Inherent operating mechanism      & Associative   & Rule-governed \\
        Adaptability  & Ability to change or evolve       & Slow-learning & Flexible      \\
    \end{tabular}
\end{table*}

\section{Propaganda Techniques}
\label{app:propaganda_techniques}
Table ~\ref{Table: Propaganda Techniques} shows the different propaganda techniques after \citet{da2019fine,abdullah2022detecting}.

\begin{table*}[!ht]
\centering
\small
\begin{tabular}{|l|p{5cm}|p{5cm}|}
\hline
\textbf{Propaganda Technique} & \textbf{Definition} & \textbf{Example} \\
\hline
Appeal\_to\_Authority & Supposes that a claim is true because a valid authority or expert on the issue supports it & "The World Health Organisation stated, the new medicine is the most effective treatment for the disease." \\ \hline
Appeal\_to\_fear-prejudice & Builds support for an idea by instilling anxiety and/or panic in the audience towards an alternative & "Stop those refugees; they are terrorists." \\ \hline
Bandwagon,Reductio\_ad\_hitlerum & Justify actions or ideas because everyone else is doing it, or reject them because it's favored by groups despised by the target audience & "Would you vote for Clinton as president? 57\% say yes." \\ \hline
Black-and-White\_Fallacy &  Gives two alternative options as the only possibilities, when actually more options exist & "You must be a Republican or Democrat" \\ \hline
Causal\_Oversimplification &  Assumes a single reason for an issue when there are multiple causes & "If France had not declared war on Germany, World War II would have never happened." \\ \hline
Doubt & Questioning the credibility of someone or something & "Is he ready to be the Mayor?" \\ \hline
Exaggeration,Minimisation & Either representing something in an excessive manner or making something seem less important than it actually is & "I was not fighting with her; we were just playing." \\ \hline
Flag-Waving & Playing on strong national feeling (or with respect to a group, e.g., race, gender, political preference) to justify or promote an action or idea & "Entering this war will make us have a better future in our country." \\ \hline
Loaded\_Language & Uses specific phrases and words that carry strong emotional impact to affect the audience & "A lone lawmaker"s childish shouting." \\ \hline
Name\_Calling,Labeling & Gives a label to the object of the propaganda campaign as either the audience hates or loves & "Bush the Lesser." \\ \hline
Repetition &  Repeats the message over and over in the article so that the audience will accept it & "Our great leader is the epitome of wisdom. Their decisions are always wise and just." \\ \hline
Slogans & A brief and striking phrase that contains labeling and stereotyping &  "Make America great again!" \\ \hline
Thought-terminating\_Cliches &  Words or phrases that discourage critical thought and useful discussion about a given topic & "It is what it is" \\ \hline
Whataboutism,Straw\_Men,Red\_Herring & Attempts to discredit an opponent's position by charging them with hypocrisy without directly disproving their argument & "They want to preserve the FBI's reputation." \\ 
\hline
\end{tabular}
\caption{Propaganda Techniques after \citet{da2019fine,abdullah2022detecting}}
\label{Table: Propaganda Techniques}
\end{table*}

\clearpage

\section{Prompts}
\subsection{Propaganda detection and explanation prompt}\label{app:propaganda-detection-prompt-template}
\begin{Verbatim}[breaklines=true,breakanywhere=true,fontsize=\small,    commandchars=\\\{\}, breaksymbol= ]
# Role
You are an expert in communication and political science and you are specialized in identifying propaganda techniques in political articles.

# Task
Your task is to carefully analyze the given article and identify any of the 14 propaganda techniques used in the text.
Always be cautious and thorough in your analysis, ensuring not to wrongly identify propaganda techniques.

## Propaganda Techniques
These are the 14 propaganda techniques you will identify, with definitions and examples (in no particular order):
- [Loaded\_Language]: Uses specific phrases and words that carry strong emotional impact to affect the audience, e.g., 'a lone lawmaker's childish shouting.'
- [Name_Calling, Labeling]: Gives a label to the object of the propaganda campaign that the audience either hates or loves, e.g., 'Bush the Lesser.'
- [Repetition]: Repeats the message over and over in the article so that the audience will accept it, e.g., 'Our great leader is the epitome of wisdom. Their decisions are always wise and just.'
- [Exaggeration, Minimization]: Either represents something in an excessive manner or makes something seem less important than it actually is, e.g., 'I was not fighting with her; we were just playing.'
- [Appeal_to_fear-prejudice]: Builds support for an idea by instilling anxiety and/or panic in the audience towards an alternative, e.g., 'stop those refugees; they are terrorists.'
- [Flag-Waving]: Playing on strong national feeling (or with respect to a group, e.g., race, gender, political preference) to justify or promote an action or idea, e.g., 'entering this war will make us have a better future in our country.'
- [Causal_Oversimplification]: Assumes a single reason for an issue when there are multiple causes, e.g., 'If France had not declared war on Germany, World War II would have never happened.'
- [Appeal_to_Authority]: Supposes that a claim is true because a valid authority or expert on the issue supports it, e.g., 'The World Health Organization stated the new medicine is the most effective treatment for the disease.'
- [Slogans]: A brief and striking phrase that contains labeling and stereotyping, e.g., 'Make America great again!'
- [Thought-terminating\_Cliches]: Words or phrases that discourage critical thought and useful discussion about a given topic, e.g., 'it is what it is'
- [Whataboutism, Straw\_Men, Red\_Herring]: Attempts to discredit an opponent's
position by charging them with hypocrisy without directly disproving their argument, e.g., 'They want to preserve the FBI's reputation.'
- [Black-and-White\_Fallacy]: Gives two alternative options as the only possibilities when actually more options exist, e.g., 'You must be a Republican or Democrat'
- [Bandwagon, Reductio\_ad\_hitlerum]: Justify actions or ideas because everyone else is doing it, or reject them because it's favored by groups despised by the target audience, e.g., 'Would you vote for Clinton as president? 57\% say yes.'
- [Doubt]: Questioning the credibility of someone or something, e.g., 'Is he ready to be the Mayor?'

# Instructions
For the given article, please identify all occurrences of the 14 propaganda techniques and provide an explanation of why each passage is considered political propaganda.
Further, return the EXACT passage where the political propaganda can be found.
If no propaganda technique was identified, return an empty dictionary.
If a propaganda technique is used multiple times in the article,, all occurrences MUST be identified and explained.

# Output Format
The output should be valid JSON, with each propaganda technique as a key and a list of occurrences as values.
Each occurrence should have an explanation and the exact passage in the article where the propaganda technique is present.

## Example:
{
    "Loaded_Language": [
        {
            "explanation": "This is an example explanation.",
            "location": "This is the exact passage in the article where the propaganda technique is present."
        },
        {
            "explanation": "This is another example explanation.",
            "location": "This is the exact passage in the article where the propaganda technique is present."
        }
    ],
    "Name_Calling, Labeling": [
        {
            "explanation": "This is another example explanation.",
            "location": "This is the exact passage in the article where the propaganda technique is present."
        }
    ]
}

Here is the article:"
\end{Verbatim}

\subsection{Contextualizer ReAct Prompt Template}\label{app:contextualizer-propmpt-template}
\begin{Verbatim}[breaklines=true,breakanywhere=true,fontsize=\small,    commandchars=\\\{\}, breaksymbol= ]
You are an Assistant tasked to contextualise potentially misleading statements to make sure users are safe and well informed. You have access to the following tools:
    Google - Get previews of the top google search results to get more information about the statement. The function always returns the next 10 results and can be called multiple times. If initial results seem unrelated you may use quotation marks to search for an exact phrase. Use a minus sign to exclude a word from the search.  Use before:date and after:date to search for results within a specific time period. Do not google the entire statement verbatim.
    
Do not use any tool more than three times. Use the following format:
    
Question: the input question you must answer
Thought: you should always think about what to do
Action: the action to take, should be one of [Google]
Action Input: the input to the action
Observation: the result of the action
... (this Thought/Action/Action Input/Observation can repeat 3 times)
Thought: I now have sufficient information to provide context for the user.
Final Answer: The context demanded by the user.

**Final Response Format:**
- **Misinformation Explanation:** (Detailed reasoning)
- **Sources:** (List all important sources with hyperlinks)
    
**Example Final Answer:**
Context: Electric vehicles are generally less harmful to the environment over their lifetime than gasoline cars, though they do have a high environmental cost at the production stage.
Sources: 
- [EPA](https://www.epa.gov/greenvehicles/electric-vehicle-myths)
- [MIT Climate Portal](https://climate.mit.edu/ask-mit/are-electric-vehicles-definitely-better-climate-gas-powered-cars)
- [NY Post](https://nypost.com/2024/03/05/business/evs-release-more-toxic-emissions-are-worse-for-the-environment-study/)

Begin!

Question: Contextualise the statement: '{statement}'{originator_section}{date_section}
Thought:{agent_scratchpad}"""
\end{Verbatim}

\section{Measurements}\label{app:measurements}

We employed a comprehensive set of measurements to assess participants' CT, propaganda awareness, and experiences with the tool. While we collected data across multiple dimensions, we focus here on the key measurements most relevant to our research questions. The complete survey included additional questions about news consumption, political orientation, and general media trust.

\paragraph{Pre-Reading Questionnaires}
To assess participants' tendency toward effortful cognitive activities, we used the Need for Cognition Scale \cite{lins_de_holanda_coelho_very_2020}. Participants rated their agreement with statements such as \textit{I would prefer complex to simple problems} on a 7-point Likert scale (1 = strongly disagree to 7 = strongly agree).
Before reading the assigned articles, participants completed questionnaires assessing their political leaning and news consumption habits.
For political leaning, participants indicated their political orientation on a 7-point Likert scale.
Regarding news consumption, participants reported the frequency with which they read news articles and rated their level of trust in the news media. They also identified the sources they typically use to consume news, selecting from a provided list and an option to add any additional sources.

\paragraph{Post-Reading Questionnaires}  
Following each reading task, participants completed several questionnaires.
To assess their critical thinking, participants were asked six question based on Kahneman.
As defined by \cite{zavolokina_think_2024} participants responded using 7-point Likert scales. Additionally, we added one open-ended question to allow us to gauge both the depth and quality of their critical thinking.
Critical to our research question, we also assessed the propaganda awareness.
To measure participants' awareness and recognition of propaganda techniques, we used two measures.
First, participants were asked to indicate whether they believed the article contained propagandistic elements and explained their reasoning in an open-ended response.
Second, participants were presented with a series of passages extracted from the articles. The passages included both statements identified as propagandistic by the tool and non-propagandistic statements. Participants were asked to determine whether each passage contained propagandistic elements (\textbf{Yes} or \textbf{No}). This task assessed their ability to recognize propaganda without the tools' assistance.

\paragraph{News Evaluation}
Participants were also asked to evaluate the article in several dimensions using 7-point Likert scales.
The dimensions were accuracy, bias, informativeness, trustworthiness, and clarity.

\paragraph{Tool Experience and Evaluation}
For participants who used the tool (Groups 3--5), additional measures were used to assess their experience and perceptions of the tool.
We included questions on the tool's usability, impact on trust and opinion, reliance, information overload, and critical thinking.
Additionally, participants were presented with three randomly selected instances of propaganda detected by the tool in \textit{Article 1}.
For each instance, they assessed the tool's accuracy, usefulness, relevance, trustworthiness, credibility, clarity and informativeness.
Participants who received the overview of propaganda techniques were asked for suggestions for improvements to the overview, how helpful it was in increasing awareness of propaganda techniques, and how likely they were to recommend the overview to others.

\subsection{Key Survey Instruments}
\label{app:survey_instruments}
\paragraph{Critical Thinking Assessment}
The following questions were used to evaluate participants' CT, based on Kahneman's dual-system thinking framework:

\begin{itemize}
    \item \textit{Speed:} "Did you feel like your thought process was more like a..." (1 = Quick skim, 7 = Slow, careful read)
    \item \textit{Processing:} "Did you feel like you were..." (1 = Absorbing several details at once, 7 = Focusing on one detail at a time)
    \item \textit{Effort:} "Did your thinking feel..." (1 = Easy and automatic, 7 = Like it required significant mental effort)
    \item \textit{Nature:} "Were you more likely to..." (1 = Make connections between details instinctively, 7 = Apply a set of rules or criteria to evaluate details)
    \item \textit{Adaptability:} "Did you find yourself..." (1 = Sticking to your initial impressions, 7 = Regularly updating your thoughts as you read more)
    \item \textit{Control:} "While reading the article, did your thoughts feel more..." (1 = Spontaneous, 7 = Deliberate)
\end{itemize}

Additionally, participants were asked an open-ended question: "Please explain why you felt this way while reading the article. Provide specific examples or reasons for your rating."

\paragraph{Propaganda Awareness Evaluation}
To assess propaganda awareness we used two measurement approaches:
\textit{Tool Impact Assessment}
Participants self-reported their experience on a 7-point Likert scale (1 = not at all, 7 = very much):
\begin{itemize}
\item "Did the tool help you become more aware of the different propaganda techniques used in media?"
\end{itemize}
\textit{Propaganda Detection Task}
Participants were presented with various statements from the articles and asked to identify whether they contain propagandistic elements or not:
\begin{itemize}
\item For each statement, participants indicated: "Does it contain propagandistic elements?" (Yes/No)
\end{itemize}

\paragraph{Feedback and Net Promoter Score}
At the end, participants who used the tool were asked to provide feedback on the tool, like suggestions for improvements, what they found most helpful, and additional comments or feedback.
We also used the Net Promoter Score, participants were asked to rate how likely they were to recommend the tool to others on a scale from 0 to 10.